\documentclass[11pt,nofootinbib]{revtex4}
\usepackage{amsfonts,amssymb,amsmath,graphicx}
\usepackage{bm}
\usepackage[spanish,english]{babel}
\usepackage{graphicx}
\usepackage[latin1]{inputenc}
\usepackage{hyperref}

\begin{document}

\newcommand{\be}{\begin{equation}}
\newcommand{\ee}{\end{equation}}
\newcommand{\bea}{\begin{eqnarray}}
\newcommand{\eea}{\end{eqnarray}}
\newcommand{\nn}{\nonumber \\}
\newcommand{\e}{\mathrm{e}}

\title{\Large Singular inflation from Born-Infeld-f(R) gravity}

\author{{\large Emilio Elizalde}
  \\ {\small Instituto de Ciencias del Espacio (ICE/CSIC)  and
Institut d'Estudis Espacials de Catalunya (IEEC) \\ Campus UAB, Carrer de Can
Magrans, s/n,
08193 Bellaterra (Barcelona), Spain} \\ }

\author{{\large Andrey N.~Makarenko}  \\ {\small
Tomsk State Pedagogical University, ul.~Kievskaya, 60, 634061 Tomsk,
Russia} \\ {\small
National Research Tomsk State University, Lenin Avenue, 36, 634050 Tomsk,
Russia}}

\begin{abstract}

Accelerating dynamics from Born-Infeld-$f(R)$ gravity are studied
in a simplified conformal approach without matter. Explicit unification of
inflation with late-time acceleration is realized within this singular inflation approach, which is similar to Odintsov-Oikonomou singular $f(R)$
inflation. Our model turns out to be consistent with the latest release of Planck data.

\end{abstract}

\pacs{11.30.-j, 98.80.Cq, 04.50.-h, 04.50.Kd}

\maketitle

\section{Introduction}

Various cosmological observations support
the current accelerated expansion of our universe.
To explain this phenomenon
it is necessary to assume either the existence of dark energy, which
has a negative pressure, or the fact that gravity must be modified.
Indeed, a quite natural approach to the universe evolution is the
description where both the early-time as well as the late-time universe acceleration is
achieved by a modification of the standard theory of General Relativity \cite{review}-\cite{review6}.
Work in this direction has shown that, indeed, a
unified description of the early-time inflation with the late-time dark energy appears
rather naturally in modified gravity, as was clearly shown by Nojiri-Odintsov \cite{R}
(for further unified models of  this sort  see Refs.~\cite{Nojiri:2007as}-\cite{N7d}). However,
the transition from a decelerating phase to the dark-energy universe is
not yet well understood (possibly because there is not yet a clear understanding of
what dark energy itself actually is).

In the recent literature, increasing interest has appeared for theories of
gravity formulated in the Palatini scheme, in particular the so-called Born-Infeld theories \cite{BI,BI2}. The Palatini formulation brings about
a number of restrictions and additional constraints to the metrics under
consideration. As a consequence, it turns out to be quite difficult to get consistent
generalizations of the original  Born-Infeld  model.

However, it was recently demonstrated that a non-perturbative and consistent
generalization of Born-Infeld gravity is possible, under the form of a new
Born-Infeld-$f(R)$ theory, which was introduced in Ref.~\cite{Eb6}. It was shown
there that  Born-Infeld-$F(R)$ gravity without matter can be easily reconstructed in
the conformal approach. In this way, eventually any kind of  dark energy cosmology could in principle be derived from the above theory.

In the present letter we will apply the techniques of \cite{Eb7}
in order to explicitly show that  Born-Infeld-$f(R)$ gravity is able to
give rise to a very realistic singular inflation theory accurately
matching the most recent Planck data.
This provides a natural possibility for the unification of singular
inflation with dark energy within the theory under discussion here.

\section{Born-Infeld-$f(R)$ theory in a conformal ansatz}

In what follows we are going to consider the Born-Infeld-$f(R)$ theory in
a conformal ansatz (see Refs.~\cite{Eb6, Eb7} for details), which can be
used with the purpose to discuss a number of relevant situations. In
particular, we can explicitly work out in detail the unification of the inflation
epoch with the late-time acceleration stage by using metrics proposed in \cite{OD1}.

As already advanced above, in order to enhance the capabilities of the theory, we
here propose a modified action of Born-Infeld type but containing an arbitrary
function $f(R)$, where
$R=g^{\mu\nu} R_{\mu\nu}(\Gamma)$ \cite{Eb6}, namely
\be
\label{act}
S=\frac{2}{\kappa}\int
d^4x\left[\sqrt{|\det{\left(g_{\mu\nu}+\kappa
R_{\mu\nu}(\Gamma)\right)}|}-\lambda\sqrt{|g|}\right]
+\int d^4x\sqrt{|g|}f(R).
\ee
As noted above, matter will be absent from our model, the purpose being to retain
the full power of the conformal approach (the massive case, with its
particularities, will be the subject of a future investigation). Under the conformal
approach we understand the situation where the  metric $g_{\mu\nu}$  and  the
auxiliary metric (on which the Christoffel symbols are built, both metrics, as is
known, being independent in the Palatini formalism) are connected by a transformation having the form of a conformal one ($g_{\mu\nu}=\Omega \, u_{\mu\nu}$).

   Varying action  (\ref{act}) with respect to the connection, the following
equation results
\be
\label{eq1}
\nabla_\alpha\left[\sqrt{q}\left( q^{-1}\right)^{\mu\nu}+\sqrt{g} g^{\mu\nu}
f_R\right]=0 \ ,
\ee
where
\be \label{q1}
q_{\mu\nu}=g_{\mu\nu}+\kappa
R_{\mu\nu}(\Gamma)
\ee
and being
$f_R\equiv df/dR$. The corresponding equation which follows by variation over the
metric has the form
\be
\label{e1_1}
\sqrt{q}\left(q^{-1}\right)^{\mu\nu}-\lambda
\sqrt{g}g^{\mu\nu}+\frac{\kappa}{2}\sqrt{g}g^{\mu\nu} f(R)-\kappa \sqrt{g}
f_R R^{\mu\nu}=0.
\ee
Since we work in the conformal approach, it is just sufficient to require the
fulfillment of the following condition
\be
\label{q1}
q_{\mu\nu}=k(t) g_{\mu\nu}.
\ee
In this case we have an auxiliary metric, $u_{\mu\nu}$,  which defines the
covariant derivative and, hence, the Christoffel symbols, as
\be
\label{metric2}
\Gamma^\alpha_{\mu\nu}=\frac{1}{2} u^{\alpha\beta}\left(\partial_\mu
u_{\nu\beta}+\partial_\nu u_{\mu\beta}-\partial_\beta u_{\mu\nu}\right),
\ee
where
\be
\label{umn}
u_{\mu\nu}=(k(t)+f_R)g_{\mu\nu}.
\ee

From the condition (\ref{q1}),  together with the definition $q_{\mu\nu}$, it is
clear that the Ricci tensor must be also proportional to the metric
$g_{\mu\nu}$. One can write the relationship between the Ricci tensor and the
metric, as
\be
\label{Ruq}
R_{\mu\nu}=\frac{1}{\kappa}[k(t)-1]g_{\mu\nu}.
\ee
Let us now consider  the spatially-flat FRW universe, with metric
\be
\label{FRW}
ds^{2}=-dt^{2}+a^{2}(t)(dx^{2}+dy^{2}+dz^{2})\  .
\ee
The auxiliary metric will be given by the expression (\ref{umn}).

We denote the function connecting the main and the auxiliary metrics by
$u(t)=k(t)+f_R$. Assume now that $R_{\mu\nu}=r(t) g_{\mu\nu}$, where $r(t)$ is easy
to find from Eq.~(\ref{Ruq}).
Finally, after all these considerations are taken into account, the equations
acquire the following form
\begin{eqnarray}
r(t)&=&3H^2+\frac{3H\dot{u}}{u(t)}+\frac{3\dot{u}(t)^2}{4u(t)^2},\\
2\dot{H}&=&H\frac{\dot{u}(t)}{u(t)}+\frac{3\dot{u}(t)^2}{2u(t)^2}-\frac{\ddot{u}(t)^2}{u(t)},
\end{eqnarray}
$H$ being the Hubble rate, $H=\frac{\dot{a}}{a}$.
From these expressions, it follows that
\be
\label{con1}
u(t)=c \,\,r(t)
\ee
where $c$ is a constant. The remaining equations lead to
\be
\label{eeq1}
H=\pm\sqrt{\frac{u}{3c}}-\frac{\dot{u}}{2u}.
\ee
From this result (see \cite{Eb6}), the form of the function $f(R)$ can be found
explicitly to be given by
\be
\label{ff1}
f(R)=\frac{2}{\kappa}(\lambda-1)-R+\frac{c-\kappa}{8}R^2.
\ee

\section{Our model}

In what follows we will go one step further and consider a particular universe with scale factor of the following type \cite{OD3,OD32}
\be
\label{eqq}
a(t)=\e^{-[f_0 (-t+t_s)^{1+\alpha }]/(1+\alpha )}\, ,
\ee
where $f_0$, $t_s$  and $\alpha$  are constants.
Interest in this type of metrics has arisen in the literature when
considering the so-called singular inflationary cosmologies \cite{OD2,
OD3,OD32}. The Hubble parameter takes the form
\be
H=
f_0 (-t+t_s)^{\alpha }
\ee
In order to look now for the relation between the metrics $g_{\mu\nu}$ and
$u_{\mu\nu}$, it is necessary to find solutions to Eq.~(\ref{eeq1}), as
\be
\label{Osts1}
u(t)=
\frac{3 (1+\alpha)^2 c e^{\frac{2 f_0 (-t+t_s)^{1+\alpha}}{1+\alpha}}
\left(-\frac{f_0
(-t+t_s)^{1+\alpha}}{1+\alpha}\right)^{\frac{2}{1+\alpha}}}{\left\{\alpha (1+\alpha)
f_0 \left(-\frac{f_0 (-t+t_s)^{1+\alpha}}{1+\alpha}\right)^{\frac{1}{1+\alpha}}
C\pm(t-t_s){\Gamma}\left[\frac{1}{1+\alpha},-\frac{f_0
(-t+t_s)^{1+\alpha}}{1+\alpha}\right]\right\}^2}.
\ee
Here $C$ is a constant and $\Gamma[a,z]$ is the incomplete gamma function.

The scale factor  $a(t)$ (\ref{eqq}) for the spatially-flat FRW universe (\ref{FRW})  is a solution of equation (\ref{e1_1}) (obtained by varying the action (\ref{act})  with respect to the connection) provided the conditions (\ref{ff1}) and (\ref{Osts1}) are imposed. In fact, taking into account
 (\ref{ff1}) and (\ref{eqq}), Eq.~(\ref{e1_1}) takes the form:
\bea
&&
-1+\Lambda-u(t)+\frac{1}{8 (t-t_s) u(t)^2}3 (c-\kappa) \left(4 f_0 (-t+t_s)^{\alpha} \left(\alpha-2 f_0 (-t+t_s)^{1+\alpha}\right) u(t)^2+\right.\nonumber\\
&&\left.+ (-t+t_s) u'(t)^2+2 (t-t_s) u(t) \left(3 f_0 (-t+t_s)^{\alpha} u'(t)+u''(t)\right)\right)+\nonumber\\
&&+\left(u(t)-\frac{1}{8 (t-t_s) u(t)^2}3 (c-\kappa) \left(4 f_0 (-t+t_s)^{\alpha} \left(\alpha-2 f_0 (-t+t_s)^{1+\alpha}\right) u(t)^2+\right.\right. \nonumber\\
&&\left.\left.+(-t+t_s) u'(t)^2+2 (t-t_s) u(t) \left(3 f_0 (-t+t_s)^{\alpha} u'(t)+u''(t)\right)\right)\right)\times\nonumber\\
&&\times \left(-1+\frac{1}{8 (t-t_s) u(t)^2}3 (c-\kappa) \left(4 f_0 (-t+t_s)^{\alpha} \left(\alpha-2 f_0 (-t+t_s)^{1+\alpha}\right) u(t)^2+\right.\right.\nonumber\\
&&\left.\left.+(-t+t_s) u'(t)^2+2 (t-t_s) u(t) \left(3 f_0 (-t+t_s)^{\alpha} u'(t)+u''(t)\right)\right)\right)-\nonumber\\
&&-\frac{1}{2} \kappa \left(\frac{2-2 \Lambda}{\kappa}-\frac{1}{2 (t-t_s) u(t)^2}3 \left(4 f_0 (-t+t_s)^{\alpha} \left(\alpha-2 f_0 (-t+t_s)^{1+\alpha}\right) u(t)^2+\right.\right.\nonumber\\
&&\left.+(-t+t_s) u'(t)^2+2 (t-t_s) u(t) \left(3 f_0 (-t+t_s)^{\alpha} u'(t)+u''(t)\right)\right)+\nonumber\\
&&+\frac{1}{32 (t-t_s)^2 u(t)^4}9 (c-\kappa) \left(4 f_0 (-t+t_s)^{\alpha} \left(\alpha-2 f_0 (-t+t_s)^{1+\alpha}\right) u(t)^2+\right.\nonumber\\
&&\left.\left.+(-t+t_s) u'(t)^2+2 (t-t_s) u(t) \left(3 f_0 (-t+t_s)^{\alpha} u'(t)+u''(t)\right)\right)^2\right)
\eea
And if we now substitute in this equation the corresponding expression for the function $u(t)$ (\ref{Osts1}) then it is not difficult to see that we obtain an identity.

We will now discuss several specific cases in more detail. First, consider the situation when $f_0<0$ ($f_0=-1$). In this case, the evolution of the
metric, of the Hubble rate, and of the equation of state (EoS) parameter are given
in Figs.~1-3, respectively.

\begin{figure}[!h]
\begin{minipage}[h]{0.25\linewidth}
\includegraphics[angle=0, width=1\textwidth]{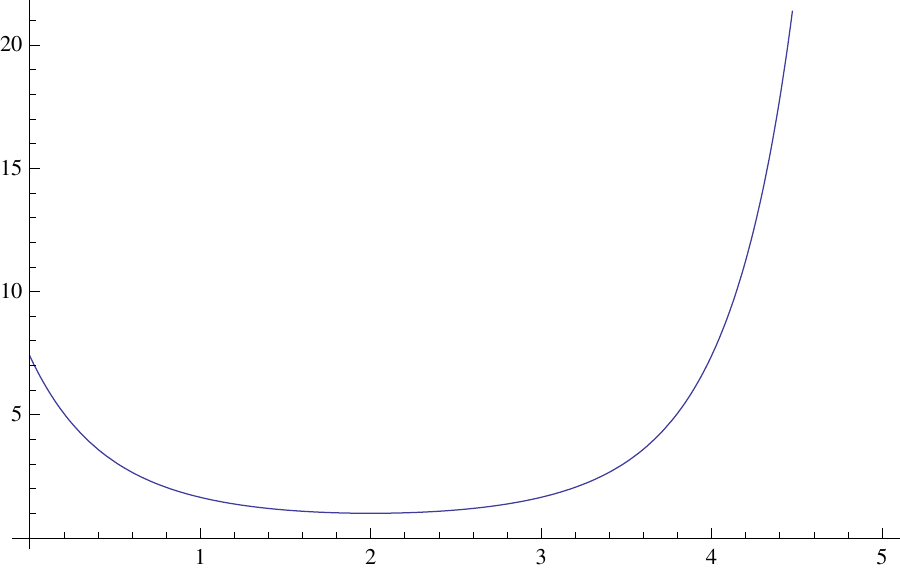}
\vspace{-5mm}
\caption{Scale factor for $f_0=-1$, $t_s=2$, $\alpha=1$}
\end{minipage}
\hspace{10mm}
\begin{minipage}[h]{0.25\linewidth}
\includegraphics[angle=0, width=1\textwidth]{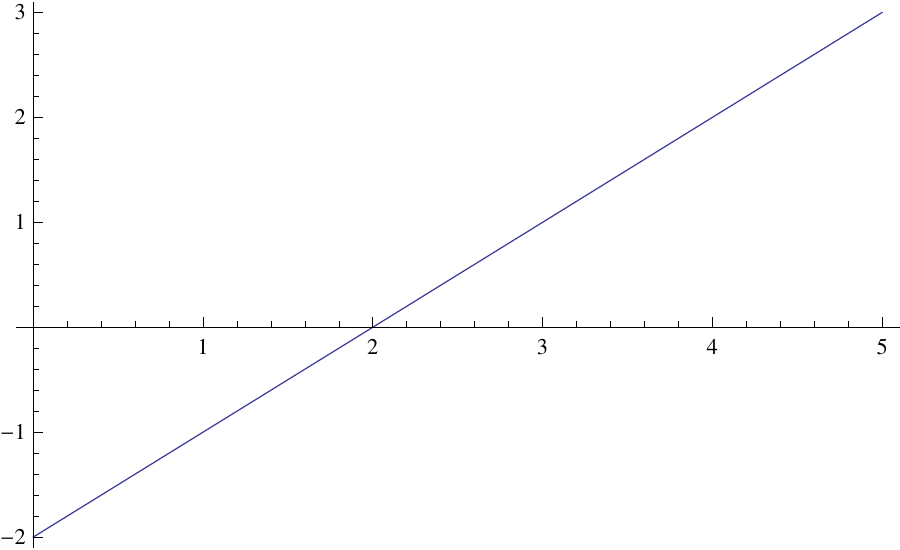}
\vspace{-5mm}
\caption{ Hubble rate for $f_0=-1$, $t_s=2$, $\alpha=1$}
\end{minipage}
\hspace{10mm}
\begin{minipage}[h]{0.25\linewidth}
\includegraphics[angle=0, width=1\textwidth]{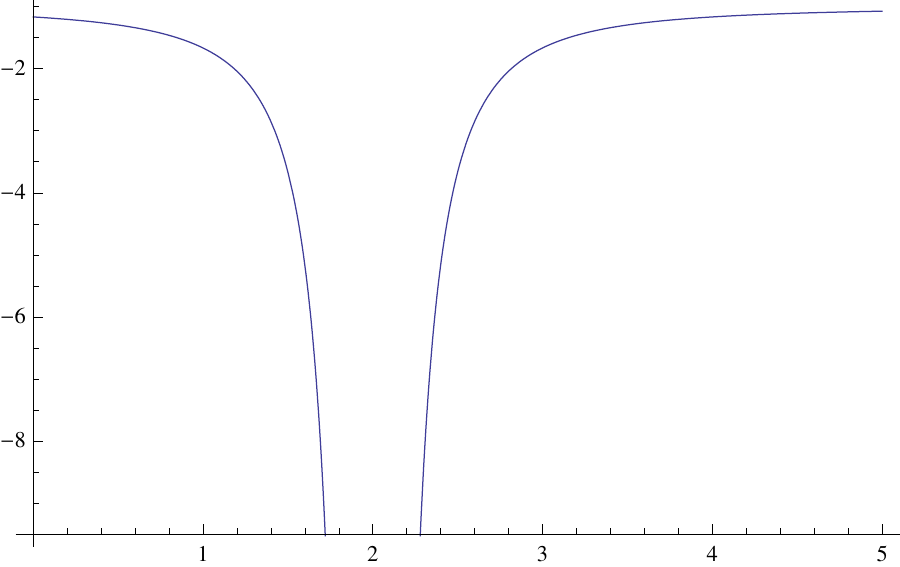}
\vspace{-5mm}
\caption{The effective EoS parameter   for $f_0=-1$, $t_s=2$, $\alpha=1$}
\end{minipage}
\label{graph_1}
\end{figure}%

For this choice of constants it is not difficult to construct the corresponding
interaction function (\ref{Osts1}), Figs.~4-5.

\begin{figure}[!h]
\begin{minipage}[h]{0.4\linewidth}
\includegraphics[angle=0, width=1\textwidth]{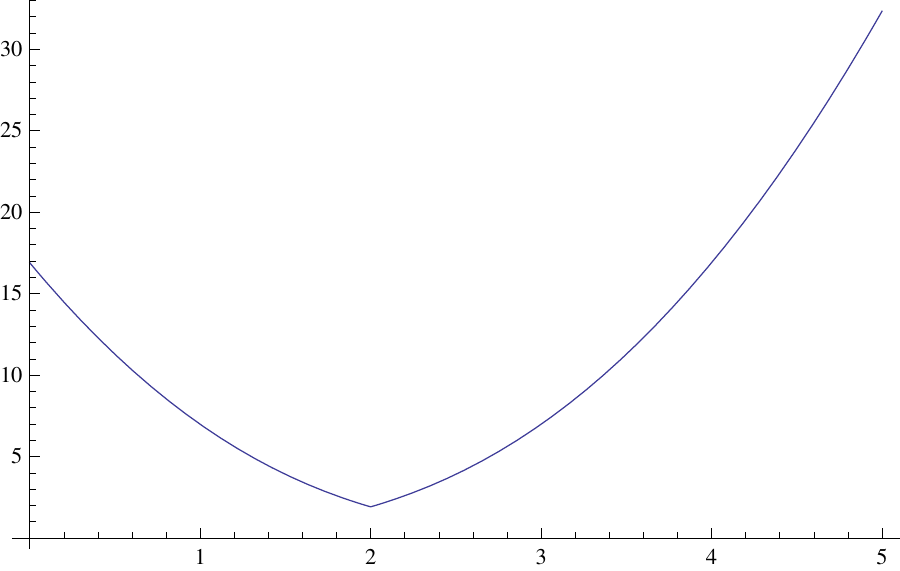}
\vspace{-5mm}
\caption{u(t) for $f_0=-1$, $t_s=2$,  $\alpha=1$, $c=1$, $C=0$}
\end{minipage}
\hspace{10mm}
\begin{minipage}[h]{0.4\linewidth}
\includegraphics[angle=0, width=1\textwidth]{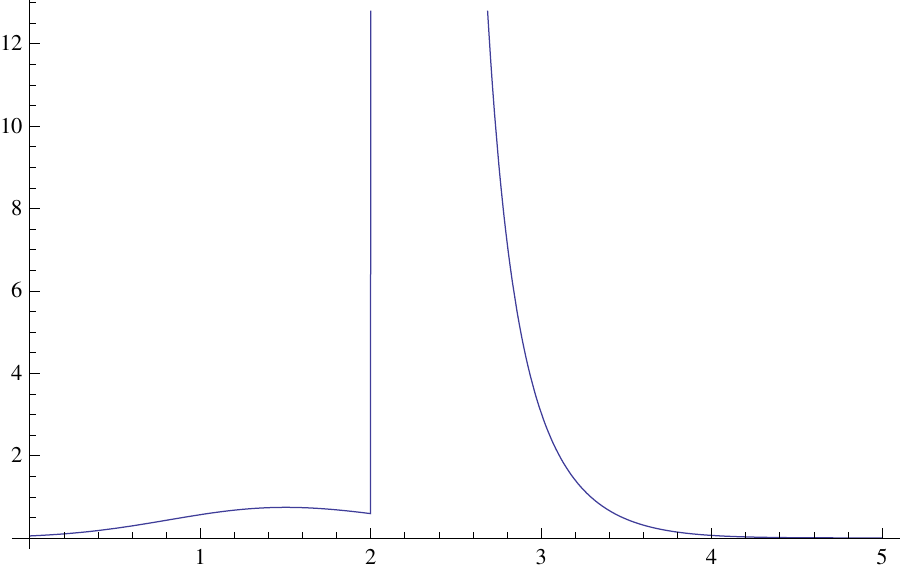}
\vspace{-5mm}
\caption{u(t) for $f_0=-1$, $t_s=2$,  $\alpha=1$,  $c=1$, $C=1$}
\end{minipage}
\label{graph_2}
\end{figure}%

One can also select alternative values for the  parameter $\alpha$. Thus, for
$\alpha=3$ one gets a different picture, as is clear from Figs.~6-7.

\begin{figure}[!h]
\begin{minipage}[h]{0.4\linewidth}
\includegraphics[angle=0, width=1\textwidth]{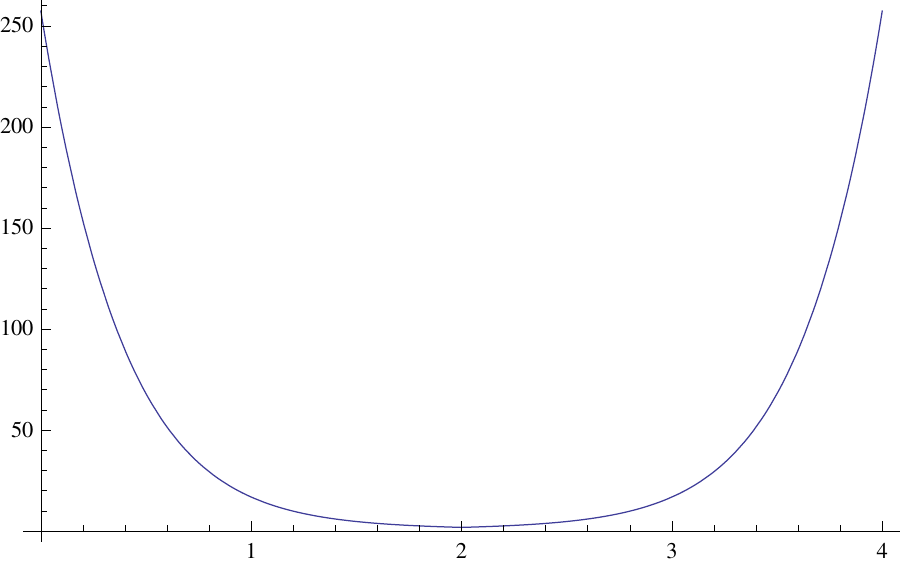}
\vspace{-5mm}
\caption{u(t) for $f_0=-1$, $t_s=2$, $\alpha=3$, $c=1$, $C=0$}
\end{minipage}
\hspace{10mm}
\begin{minipage}[h]{0.4\linewidth}
\includegraphics[angle=0, width=1\textwidth]{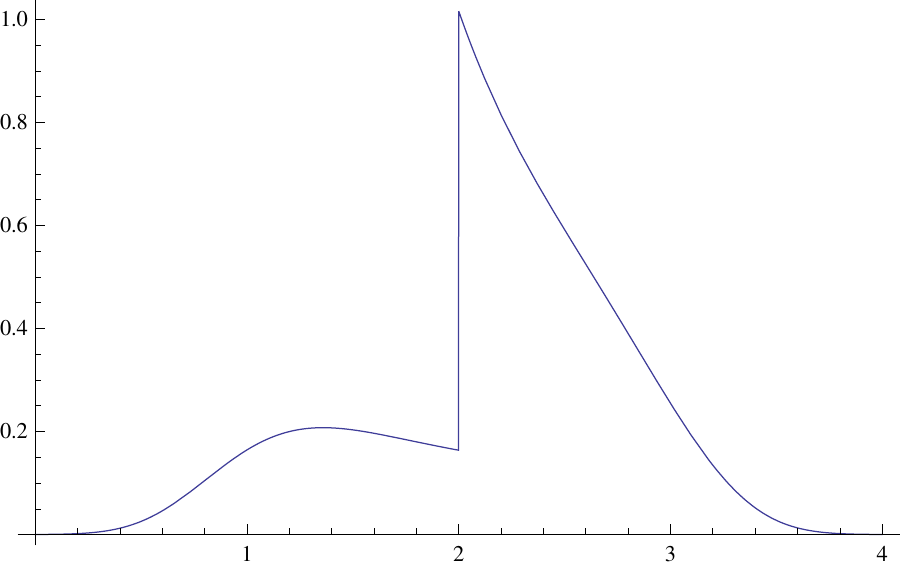}
\vspace{-5mm}
\caption{u(t) for $f_0=-1$, $t_s=2$,  $\alpha=3$,  $c=1$, $C=1$}
\end{minipage}
\label{graph_3}
\end{figure}

For even values of the parameter $\alpha$ we are bound to have  half of the
solutions only. Thus, for $\alpha=2$, we end up with Figs.~8-9.

\begin{figure}[!h]
\begin{minipage}[h]{0.4\linewidth}
\includegraphics[angle=0, width=1\textwidth]{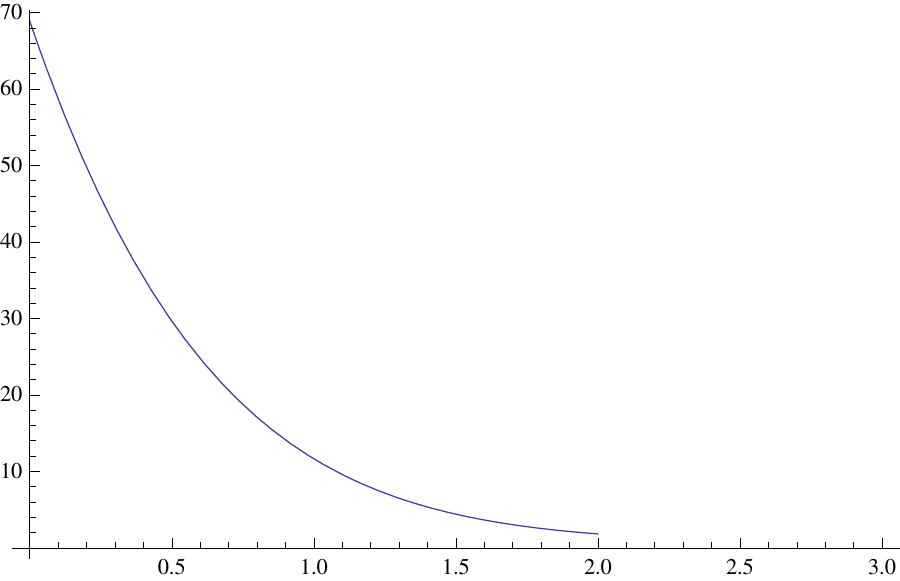}
\vspace{-5mm}
\caption{u(t) for $f_0=-1$, $t_s=2$,  $\alpha=2$, $c=1$, $C=0$}
\end{minipage}
\hspace{10mm}
\begin{minipage}[h]{0.4\linewidth}
\includegraphics[angle=0, width=1\textwidth]{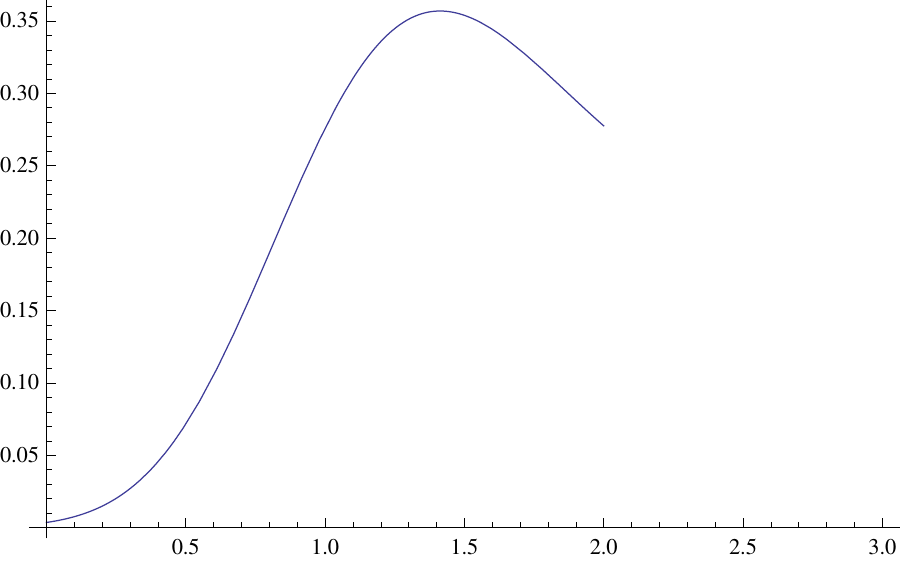}
\vspace{-5mm}
\caption{u(t) for $f_0=-1$, $t_s=2$,  $\alpha=2$,  $c=1$, $C=1$}
\end{minipage}
\label{graph_4}
\end{figure}

The opposite situation occurs for positive values of the parameter  $f_0$.  For
example, Figs.~10-13 depict the the behavior of
the metric, the Hubble rate and the EoS parameter, respectively, for $f_0=1$,
$t_s=2$, and $\alpha=2$.

\begin{figure}[!h]
\begin{minipage}[h]{0.25\linewidth}
\includegraphics[angle=0, width=1\textwidth]{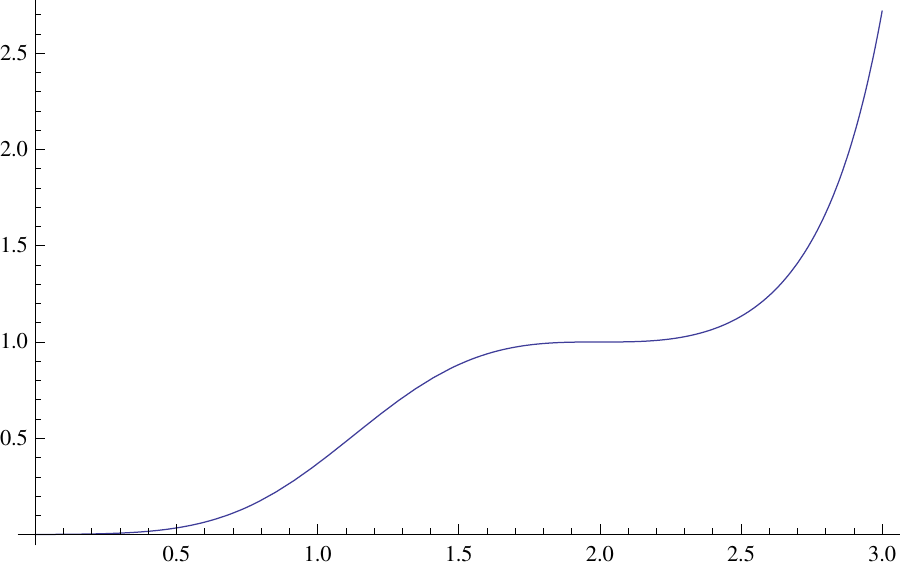}
\vspace{-5mm}
\caption{Scale factor for $f_0=1$, $t_s=2$, $\alpha=2$}
\end{minipage}
\hspace{10mm}
\begin{minipage}[h]{0.25\linewidth}
\includegraphics[angle=0, width=1\textwidth]{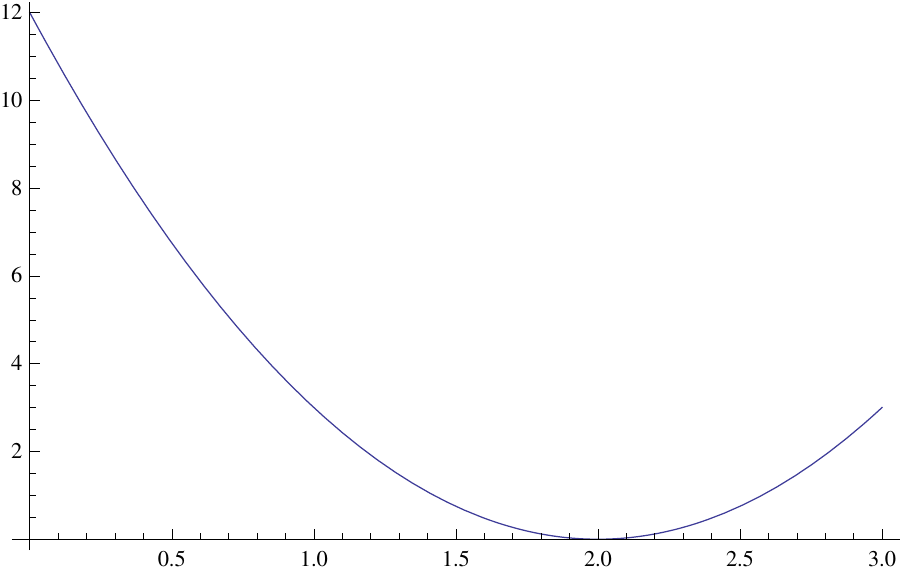}
\vspace{-5mm}
\caption{ Hubble rate for $f_0=1$, $t_s=2$, $\alpha=2$}
\end{minipage}
\hspace{10mm}
\begin{minipage}[h]{0.25\linewidth}
\includegraphics[angle=0, width=1\textwidth]{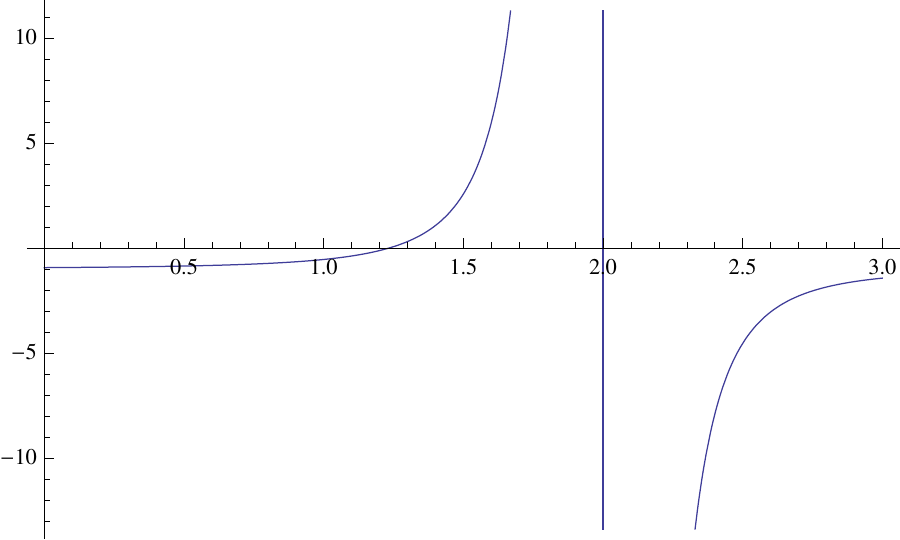}
\vspace{-5mm}
\caption{The effective  EoS parameter  for $f_0=1$, $t_s=2$, $\alpha=2$}
\end{minipage}
\label{graph_1}
\end{figure}%

In this case, the function connecting  the metrics  (\ref{Osts1}) only exists for
$t>t_s$ or for $t<t_s$, either.

\begin{figure}[!h]
\begin{minipage}[h]{0.4\linewidth}
\includegraphics[angle=0, width=1\textwidth]{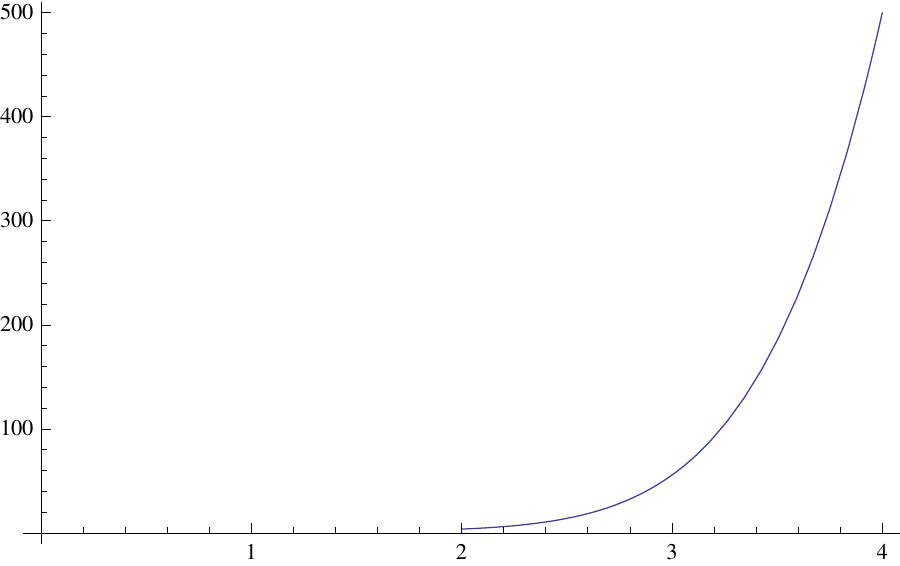}
\vspace{-5mm}
\caption{u(t) for $f_0=3$, $t_s=2$,  $\alpha=2$, $c=1$, $C=0$}
\end{minipage}
\hspace{10mm}
\begin{minipage}[h]{0.4\linewidth}
\includegraphics[angle=0, width=1\textwidth]{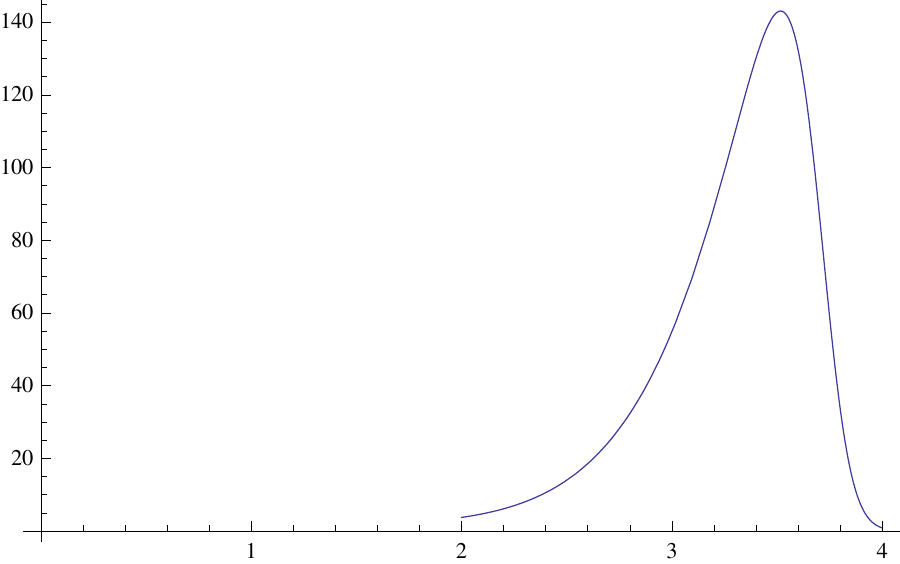}
\vspace{-5mm}
\caption{u(t) for $f_0=3$, $t_s=2$, $\alpha=2$,  $c=1$, $C=0.0001$}
\end{minipage}
\label{graph_3}
\end{figure}

To summarize, we have here demonstrated that for $\alpha$ bigger than 1, when the so-called Type  IV singularity \cite{OD4}  occurs,  we get singular inflation of the sort observed in Ref.~10. Then, precisely in the same way as in Ref.~10 one can demonstrate that if a Type IV singularity occurs at the end of inflation it may naturally induce a graceful exit from it (see  \cite{OD5})

  \section{Conclusions}

After reviewing the non-perturbative, consistent generalization of
Born-Infeld gravity, under the form of the so-called Born-Infeld-$f(R)$ theory,
introduced recently \cite{Eb6}, we have here continued this development by
applying the techniques of \cite{Eb7} to explicitly show, with specific
examples, that  Born-Infeld-$f(R)$ gravity is able to give rise to very
interesting universe models and to realize singular inflation.

For a metric chosen in the form (\ref{eqq}) and with the same constants of the theory as in Ref.~\cite{OD6}, we have obtained for the inflationary parameters,
calculated in a similar way, the following values: $n_s = 0.96491,\;\;\; r = 0.1403,\;\;\;a_s =-0.000307$. These values of the inflationary parameters are in agreement with the observational bounds for them obtained from the very latest release of Planck data.

To finish, here the Born-Infeld-$f(R)$ theory has been discussed in a conformal ansatz, without matter, with the purpose to retain the full power of the conformal approach. The massive case, with its particularities, will be the subject of a future
investigation.

\medskip

\noindent {\bf Acknowledgements}. E.E. was supported in part by MINECO (Spain),
Project FIS2013-44881-P, by the CSIC I-LINK1019 Project, and by the CPAN Consolider Ingenio Project.
A.N.M. was supported by a grant of the Russian Ministry of Education and Science.


\begin{thebibliography}{99}

\bibitem{review}
S. Nojiri and S.~D.~Odintsov,
    eConf C {\bf 0602061} (2006) 06
[Int.\ J.\ Geom.\ Meth.\ Mod.\ Phys.\  {\bf 4} (2007) 115] [ hep-th/0601213].

\bibitem{review2}
S. Nojiri and S.~D.~Odintsov,
Phys.\ Rept.\  {\bf 505} (2011) 59  [arXiv:1011.0544 [gr-qc]]. 

\bibitem{review3}
S. Nojiri and S.~D.~Odintsov, Int.\ J.\ Geom.\
Meth.\ Mod.\ Phys.\  {\bf 11} (2014) 1460006
    [arXiv:1306.4426 [gr-qc]]. 
    
    \bibitem{review4}
V.~Faraoni and S.~Capozziello,
``Beyond Einstein gravity : A Survey of gravitational theories for cosmology
and astrophysics,''
Fundamental Theories of Physics, Vol. 170, Springer, 2010. 

\bibitem{review5}
S.~Capozziello and
M.~De Laurentis,
    Phys.\ Rept.\  {\bf 509} (2011) 167
    [arXiv:1108.6266 [gr-qc]]. 
    
    \bibitem{review6}  G.~J.~Olmo,
   Int.\ J.\ Mod.\ Phys.\ D {\bf 20}, 413 (2011)
   [arXiv:1101.3864 [gr-qc]].


\bibitem{R}
   S.~Nojiri and S.~D.~Odintsov,
    Phys.\ Rev.\ D {\bf 68} (2003) 123512
    [hep-th/0307288].

\bibitem{BI}
S. Deser and G. W. Gibbons, Class. Quant. Grav. {\bf 15} (1998) L35.

 \bibitem{BI2}
M. Ba\~nados and P. G. Ferreira, Phys. Rev. Lett. {\bf 105} (2010) 011101.


\bibitem{Nojiri:2007as}
  S.~Nojiri and S.~D.~Odintsov,
   Phys.\ Lett.\ B {\bf 657} (2007) 238
[arXiv:0707.1941 [hep-th]].
    
\bibitem{N7b}  S.~Nojiri and S.~D.~Odintsov,
  Phys.\ Rev.\ D {\bf 77} (2008) 026007
[arXiv:0710.1738 [hep-th]].

\bibitem{N7bb}  E.~Elizalde, J.~E.~Lidsey, S.~Nojiri and S.~D.~Odintsov,
Phys.\ Lett.\ B {\bf 574} (2003) 1.
   
   \bibitem{N7c} G .~Cognola, E.~Elizalde, S.~Nojiri, S.~D.~Odintsov, L.~Sebastiani and
  S.~Zerbini,
Phys.\ Rev.\ D {\bf 77} (2008) 046009
[arXiv:0712.4017 [hep-th]].
   
   \bibitem{N7d}   E.~Elizalde, S.~Nojiri, S.~D.~Odintsov, L.~Sebastiani and S.~Zerbini,
Phys.\ Rev.\ D {\bf 83} (2011) 086006
  [arXiv:1012.2280 [hep-th]].


\bibitem{Eb6}
A.N. Makarenko, S. Odintsov, G.J. Olmo,
Phys.Rev. D {\bf 90} (2014) 024066
[arXiv:1403.7409 [hep-th]].

\bibitem{Eb7}
A.N. Makarenko, S. Odintsov, G.J. Olmo,
Phys.Lett. B {\bf 734} (2014) 36
[arXiv:1404.2850 [gr-qc]].

\bibitem{Eb67} A.N. Makarenko,
Astrophys.Space Sci. {\bf 352} (2014) 921-924
[arXiv:1406.1705 [gr-qc]].

\bibitem{OD1}
  S.~Nojiri and S.~D.~Odintsov,
Gen.Rel.Grav. {\bf 38} (2006) 1285
      [hep-th/0506212].

\bibitem{OD2}
S.~D.~Odintsov, V.~K.~Oikonomou,
  Phys.Rev. D{\bf 92} (2015) 2, 024016
       [arXiv:1504.06866 [gr-qc]].

\bibitem{OD3}
   S.~Nojiri,  S.~D.~Odintsov, V.~K.~Oikonomou,
Phys.Rev. D {\bf 91} (2015) 8, 084059
       [arXiv:1502.07005 [gr-qc]]. 
       
       \bibitem{OD32}
       S.~Nojiri,  S.~D.~Odintsov, V.~K.~Oikonomou,
Phys.Lett. B747 (2015) 310-320
[arXiv:1506.03307 [gr-qc]].

\bibitem{OD4}
 S.~Nojiri, S.~D.~Odintsov and S.~Tsujikawa,  Phys.\ Rev.\ D {\bf 71} (2005) 063004    [hep-th/0501025].

\bibitem{OD5}
 S.~D.~Odintsov and V.~K.~Oikonomou,
   Phys.\ Rev.\ D {\bf 92} (2015)  024058
   [arXiv:1507.05273 [gr-qc]]

\bibitem{OD6}
S.~Nojiri,  S.~D.~Odintsov, V.~K.~Oikonomou, E.~N.~Saridakis,
JCAP 1509 (2015) 044
[arXiv:1503.08443 [gr-qc]].

\end{thebibliography}
\end{document}